# EVALUATION OF A BLOCKCHAIN-ENABLED RESOURCE MANAGEMENT MECHANISM FOR NGNS

Michael Xevgenis [1,*], Dimitrios G. Kogias [2], Ioannis Christidis [1], Charalampos Patrikakis [2] and Helen C. Leligou [1]

[1]Dept. of Industrial Design and Production Engineering, University of West Attica, 122 43 Attica, Greece; auto45056@uniwa.gr, e.leligkou@uniwa.gr
[2]Dept. of Electrical and Electronics Engineering, University of West Attica, 122 43 Attica, Greece; dimikog@uniwa.gr, bpatr@uniwa.gr
*Correspondence: mxevgenis@uniwa.gr

## ABSTRACT

*A new era in ICT has begun with the evolution of Next Generation Networks (NGNs) and the development of human-centric applications. Ultra-low latency, high throughput and high availability are few of the main characteristics of modern networks. Network Providers (NPs) are responsible for the development and maintenance of network infrastructures ready to support the most demanding applications that should be available not only in urban areas, but in every corner of the earth. The NPs must collaborate to offer high quality services and keep their overall cost low. The collaboration among competitive entities can in principle be regulated by a trusted 3rd party or by a distributed approach/technology which can guarantee integrity, security, and trust. This paper examines the use of blockchain technology for resource management and negotiation among NPs and presents results of experiments conducted in a dedicated real testbed. The implementation of the resource management mechanism is described in a Smart Contract (SC), and the testbeds use the Raft and the IBFT consensus mechanisms respectively. The goal of this paper is two-fold: to assess its performance in terms of transaction throughput and latency so that we can assess the granularity at which this solution can operate (e.g. support resource re-allocation among NPs on micro-service level or not) and to define implementation-specific parameters like the consensus mechanism that is the most suitable for this use case based on performance metrics.*

## KEYWORDS

*Blockchain, NGNs, Resource management, Consensus, Performance*

## 1. INTRODUCTION

A new era in the field of computer networks has begun with the evolution towards 5G and beyond. The expansion of internet services across the world is boosted by the growth of Internet of Things (IoT) technology, mobile and wearable devices. According to Cisco's annual report, Internet users are expected to reach the number of 5,3 billion by the year of 2023 while the number of Internet connected IoT and end-user devices will reach approximately the 29,3 billion [1]. This report highlights the need for a powerful network that can handle the continuously increasing load and offer the desired Quality of Service (QoS) to customers to support intensive applications on-demand.

Nowadays, the spread of 5G networks across the world proceeds in an extremely rapid pace. The majority of Network Providers (NPs) invests in hardware in order to upgrade their existing infrastructure and deploy new Points of Presence (PoP) around the world. 5G networks aim to provide features such as, ultra-low latency, high throughput and high availability which are mandatory for the proper functionality of newly developed applications. Artificial Intelligence applications, IoT ecosystems and Machine to Machine (M2M) communication, are only few use cases where the characteristics of 5G play a vital role. Aside to 5G networks and the proliferating optical networks that come continuously closer to the user, the virtualization of network services comes to facilitate the efficient utilization of the available infrastructures.

While the hardware is used to achieve the desired propagation characteristics, the virtualization frameworks offer efficient management of network resources and programmable network capabilities. Beyond the achievements made in hardware components, the use of virtualization technology was crucial. Virtualization offered an abstraction layer above the hardware and allowed network engineers to create and manage multiple network slices, i.e., an isolated set of network resources configured to serve specific applications. In addition, the advent of virtualization led to the development of programable networks such as Software Defined Networks (SDN) and Virtual Network Functions (VNFs), as well as to the support of network services using resources offered by cloud environments. As it is described by the 5GPPP architecture working group in their whitepaper [2], new stakeholder roles have emerged in the 5G ecosystem, such as the Virtualization Infrastructure Service Provider (VISP) who interacts with the Network Operator (NOP). VISP is responsible to provide virtualized infrastructure services to NOPs while the NOPs are responsible for the management and creation of network services supported by one or multiple VISPs [3].

The efficient management of virtualized network resources requires the existence of frameworks which can interoperate among different cloud environments. In NGNs, the network resources are managed and orchestrated by MANO (Management and Orchestration) platforms. MANOs are used for the management of network resources, such as network slices and VNFs, that use resources offered by clouds for their deployment. These platforms can be connected to more than one cloud infrastructure which may belong to different owners and, therefore, they support multi-administrative scenarios. Many MANOs have been developed recently such as OSM MANO, ONAP, SONATA [4,5]. The most popular is the OSM MANO which interacts with clouds registered as Virtual Infrastructure Managers (VIMs). When resources must be used for the instantiation of network services, MANO interacts with the VIMs and deploys the requested functions. Network functions are hosted in Virtual Machines (VMs), which are described by a collection of computational resources (CPU, memory, storage). The use of MANOs and cloud computing environments for the support of network services, adds extra flexibility and scalability to NGNs and lead to new research topics.

At the same time, the characteristics of NGNs automatically force NPs to reshape their business strategy and invest in new hardware and technologies (i.e., virtualized infrastructures). The demand for high network standards worldwide increases the CAPEX and OPEX of NPs, which must collaborate in order to increase their profits and reduce their costs. Their collaboration should be based on the efficient use of hardware and virtualized resources with respect to the QoS provided to customers. The idea is that each NP offers available (unutilized) resources which can be used from another NP for a specified period of time for the agreed price. The NP that offers resources (NP offeror) should agree with the NP that request resources (NP requestor) on a Service Level Agreement (SLA) to facilitate the proper operation of the service. Since NPs invest in resources (physical or virtual) they should produce profit whether their resources are used by their own customers or by other NPs. The goal for each NP in the ecosystem is to increase the utilization of resources and, as a result increase the revenues. This approach requires the formation of a marketplace where NPs trade resources by performing transactions. Major challenges in this concept are the establishment of trust in this competitive market, the integrity of the process and the absence of a central point of control as it introduces the Single Point of Failure (SPoF) problem. Moreover, the process of lending resources among NPs must be completed in minutes or even seconds as high-quality network services must support user's applications on demand. The latency factor is critical and must be studied exhaustively as it is tightly coupled with the validity of the concept of such an approach.

A technology that answers to these challenges is blockchain, which is one of the current trends in the ICT world. Blockchain establishes trust in trustless environments while it forms a decentralized network of peers, called nodes. The decentralized feature eliminates the SPoF problem, as the network consists of many peers which contribute to the proper functionality of the system. Blockchain inherently ensures the integrity and security of transactions among users. The validated transactions of the system are recorded into blocks linked one another. The next block contains the hash of the previous block and therefore the integrity of the information stored is guaranteed. The blocks are written in a digital ledger maintained in every node of the network. Initially, blockchain was developed to perform digital currency (or cryptocurrency) transactions with its most known application the Bitcoin. Nevertheless, the last decade blockchain technology presented major advancements and is used not only in cryptocurrency transactions but also in the supply chain sector [6], in the IoT ecosystems (IOTA) [7], and in gaming [8] among others. The use of Smart Contracts (SCs) added more functionality in the blockchain technology without affecting the high security level. SCs are practically pieces of code available in the network which can be triggered by entities in the blockchain. Many blockchain platforms take advantage of SCs' functionalities such as Ethereum, Hyperledger Fabric, Corda, and others.

Due to the hype of blockchain and the evolution of NGNs, the idea of combining these two technologies attracted the interest of academia and industry, as will be thoroughly described in section 2. However, none of the proposed approaches address the use of blockchain for the management of network resources in a multitenant environment, which is the problem we focus and shed light on. In the current work, we propose a blockchain-enabled "marketplace" for the resources of the different NPs. We: a) propose a Smart Contract which can be used to perform dynamic resource management, b) highlight the requirements for the proper operation of the blockchain-based NPs' marketplace and, c) examine the behaviour of the system when different consensus mechanisms are implemented (i.e., Raft and the Istanbul Byzantine Fault Tolerant-IBFT) and select the most suitable one, d) test the performance of this solution. In contrast to the surveyed studies, we perform experiments on real testbeds that emulate geographically staggered private networks, in order to test the behaviour of this solution in terms of throughput (TPS), latency and success rate.

To this aim, this paper is organized as follows: in Section 2 we survey existing research works related to the use of blockchain for resource management in NGNs. In Section 3 we dive deeper into the requirements of this solution, the necessary functionality and the smart contract that describes the operation of the marketplace consisted of NPs and we present the main characteristics of the consensus algorithms tested. Section 4 describes the methodology of our experiment, the tools we used and the testbeds, followed by the evaluation of the collected results. The tool used for our experiments was the Hyperledger Caliper, which was initially presented in [9]. This tool is able to test the smart contract in different testbeds and extract information regarding the latency introduced by the blockchain networks. Finally, Section 5 concludes this paper.

## 2. RELATED WORK

The number of research papers studying and proposing the use of blockchain for resource management in NGNs is increasing and resulting in many interesting works to be published lately. In [10], Togou et al. present a distributed blockchain-based broker (DBB) for the dynamic leasing of resources among different network operators to support end-to-end services in a multi-administrative network. DBB includes a biding mechanism used for the management of incoming request and the construction of Memorandums of Understanding (MoUs) among operators. This solution guarantees that SLAs among operators are fulfilled. The biding process requires the proposition of bids by the operators, which are inserted into the blockchain as transactions. Then the operator who requests resources selects the cheapest one. However, the insertion of all proposed bids and not only the wining ones in the blockchain arise scalability issues. Also, the introduction of auctioning can lead to time variations, while the requests of resources must be served as soon as possible. The monitoring of the leased resources is based on a QoS matrix that is not included in the blockchain, which means that it is not fully protected from a malicious operator who may try to cheat. Moreover, the experimental part of this approach consists of a simulation which does not take into account the impact of blockchain on the performance of the system.

Maksymyuk et al. in [11] discuss the potential benefits and challenges of the integration of blockchain in the mobile network infrastructure in terms of spectrum and infrastructure sharing. Authors propose a blockchain-based framework for decentralized 6G mobile networks to ensure cooperative network management by multiple Mobile Network Operators (MNOs). Due to the potentially huge number of transactions per second produced by mobile networks, the performance of the framework is very sensitive to the "speed" of the blockchain network. The speed of the network is influenced by the underlying consensus algorithms. Therefore, this work presents the use of a new consensus, the Proof of Formulation (PoF) used in FLETA blockchain which according to authors can reach more than 10,000 transaction per second. They propose a combination of permissionless (public) and permissioned (consortium) blockchains. However, the authors have not clearly indicated which platform they have used to achieve these results and do not give the evaluation details.

In [12], Xu et al. present use cases of blockchain in next generation networks in a very abstract manner. Focusing on network slicing and resource management, authors propose the use of blockchain and SCs to introduce transparency and fairness to the system. The trading of a network slice is based on blockchain, where the SC orders the slice orchestration based on the agreed SLA described in the 5G network slice broker. The blockchain is integrated to store the usage of each leased resource and check the performance of a service provider against the SLA. According to authors, the key benefit that is introduced through the blockchain is the establishment of a trust layer, which lowers the collaboration/cooperation barrier and enables an effective and efficient ecosystem. Also, blockchain prevents the SPoF problem and thus improves systems' security. Moreover, one of the elements that play a significant role according to authors to the performance of this solution is the consensus mechanism. However, they stay in a theoretical level.

Papadakis et al. propose in [13] a blockchain-based Network Service Marketplace (NSM) and a resource orchestration mechanism that enables the Cross Service Communication (CSC) in edge cloud (EC) for the creation of NSM. Authors present a complete solution for a multi-tenant edge cloud ecosystem described by an architectural diagram. The main functionalities of the NSM are the registration, the advertisement, the discovery, the lease, the usage, and the billing. In registration phase, the tenant of an EC enters the solution and offers its services which are advertised in the network. In the discovery phase the users can browse and select the desired services and proceed to the lease. The usage of the services is monitored in order to perform the billing at the end of the lease. The blockchain layer handles through SCs all information required regarding users, services etc. Authors select the Hyperledger Fabric platform for the implementation of their solution and conduct experiments to test the performance regarding the transaction per second (TPS) and latency of the transactions implemented in the blockchain network. However, the tested network is deployed on a single VM which means that the impact on nodes communication through the internet (e.g. introduction of latency) is not taken into consideration.

Hewa et al. in [14] present the role of blockchain in 6G networks and the benefits introduced by this technology such as, privacy, integrity, and accountability. The authors focus on the application areas of this technology in 6G systems, as for example, in industrial application for beyond industry 4.0, smart healthcare, decentralized and seamless environmental monitoring and protection. Also, this paper discusses the use of blockchain for achieving decentralized network management to achieve better resource management, enhance the SLA management and spectrum sharing. In [15] Praveen et al., describe the idea of a blockchain-enabled slice broker and how the use of SCs can leverage the negotiation process among NPs in terms of automation and security. Also, this work examines the use of blockchain technology in spectrum allocation, sharing and management by the implementation of Dynamic Spectrum Sharing. Furthermore, the interest of academia and industry for the combination of blockchain and NGNs can be proven by the participation of companies such as, Intracom and Atos, to research projects like 5G Zorro [16]. The main concept of this project is to use blockchain SCs for network and security management.

The current work aims to present the evaluation of a blockchain based solution for the secure resource management in NGNs. None of the works available in the literature have shed light on the use of different consensus algorithms. Performance is evaluated on the reported articles under the assumption of co-located nodes, which is not the case in real-life. In contrast to other works, this article examines the performance of the blockchain system in terms of transaction latency, transaction throughput and success rate by conducting experiments on a real testbed that consists of geographically staggered nodes. Finally, the results of the experiment illustrate the feasibility of the solution and the improvements that should be made.

## 3. THE PROPOSED SOLUTION AND THE DESIGN ALTERNATIVES

The implementation of the blockchain enabled mechanism for resource management in NGNs consists of the blockchain platform and the pure networking part. These two parts interoperate to take advantage of the blockchain's major benefits. The proposed solution is illustrated in Figure 1. The marketplace consists of several NPs, each operating a blockchain node and a MANO instance. MANO is responsible to orchestrate the use of its own resources that consist of one or more NFV Infrastructures (NFVIs). In each one of the MANO instances, the corresponding monitoring component is aware of the level of resource utilization, as well as of the quality of service experienced by the deployed services. Currently, this component can trigger the re-configuration of the resources that the specific MANO administers, including "leased" resources which are statically defined upon agreement. In Figure 1, a solution where each area (using a MANO instance) is maintaining a blockchain node (BCi) is shown. Each BCi shown in the figure is assumed to host the wallet of the blockchain solution and the blockchain node that contains the digital ledger. This entity is triggered by the MANO when a need for additional resources occurs or when available resources are detected by the monitoring component. The blockchain entity communicates with the MANO components via an oracle service which acts as a gateway of the BCi to the outside world. Due to the fact that oracles leverage the functionality of blockchain solutions, their popularity increases and can be used to interact with other web services.

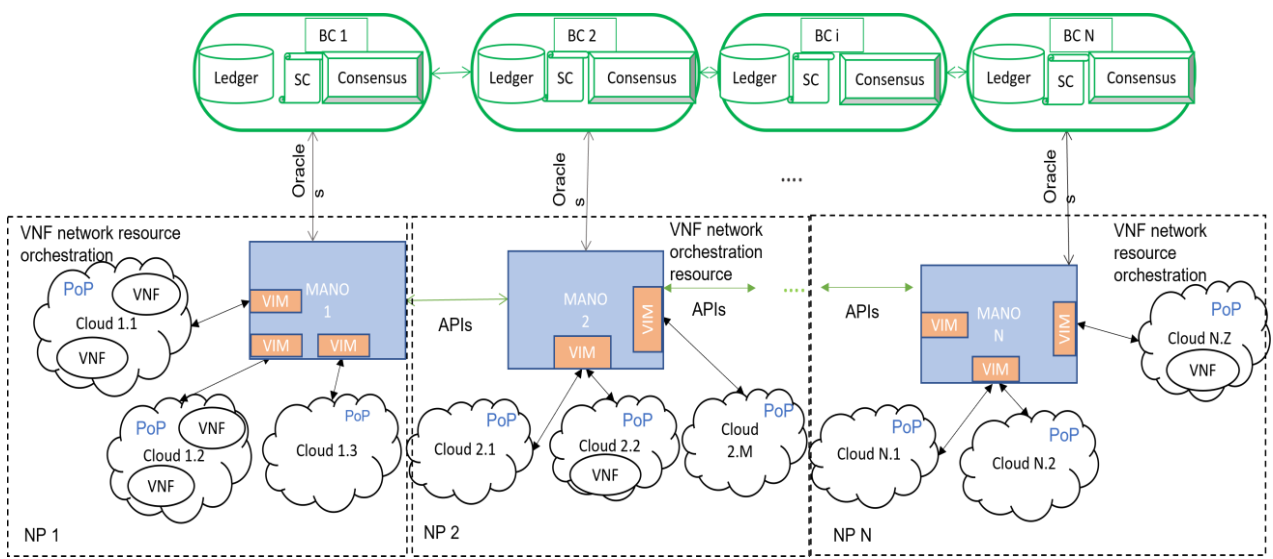

Figure 1. Architecture of the blockchain- enabled resource management solution

## 3.1. Analysis of the SC functionality

The blockchain enabled resource management mechanism is implemented through a SC that consists of functions written in Solidity [17], which is a language used for the creation of SC in Ethereum-based blockchains. The presented SC is deployed in a Quorum network which is a variation of Ethereum [18]. Quorum is ideal for the creation of private networks and supports various consensus algorithms, which is the main reason its selection. There are three main functions in this solution: addNetworkProvider, requestResources, returnResources. It is worth mentioning that every function writes in the digital ledger of the blockchain. The role of each function is described as follows:

- addNetworkProvider: This function is triggered by the administrator entity of the system to insert a new Network Provider (NP) to the blockchain. The NP is described by the following features: a) Name: the name of the NP, b)Computational resources: these resources consist of the amount of CPU, RAM, and storage the NP offers which change over time based on their utilization; c) Cost: the cost of the resources offered by the NP defined as the cost per resource; d) Domain: the area where the NP can offer the resources,; e) SLAs: the Service Level Agreements (SLAs) a provider can guarantee; f) VNF images: the Virtual Network Function (VNFs) a provider can support; g) Address: the blockchain address associated to the NP, which is used for implementing transactions in the blockchain network. The SLA describes the requirements that should be met when the resources are offered. More specific, characteristics such as latency, throughput and packet loss tolerance are defined in this field.

- requestResources: The NP who needs resources triggers this function that searches the ledger to find the NP who meets certain criteria. The criteria are based on the features analyzed above, while the requester sets the desired values of these properties. Moreover, this function uses another variable to set the time of using the resources. Summarizing, the request of resources contains the following attributes: a) Computational resources: amount of CPU, RAM, Storage; b) Domain; c) SLA; d) VNF image and e) Lend time. The execution of this request may return more than one results. In that case, the cheapest NP is selected based on the cost value. Then a transaction is initiated among the NP who called this function (requester) and the selected NP (supplier), where the supplier lends the defined number of resources to the requester for a specified time period. The requester prepays the supplier based on a cost function that takes into account the total amount of resources lent, the lend time period and the cost value. It is worth mentioning that this function includes mechanisms to ensure that the requester has the necessary balance to execute this transaction. If the requester does not have enough balance, then the transaction is reverted.

- returnResources: this function is executed when the predefined time has passed and is responsible for returning the lent resources to the original owner. This function includes an oracle mechanism to check the time and then it uses the values contained in the request transaction to return the correct number of resources to the original owner. The properties used are: a) Supplier ID: is the id of the provider who offered the resources and corresponds to a blockchain address assigned to this particular NP; b) Computational resources: amount of CPU, RAM, Storage and c) time: the time when the requestResources function was validated, which is used to check if the time has passed or not. If the time has not passed the transaction is reverted.

```
  function addNetworkProvider( _name, _cpu, _ram, _storage, _cost, _domain, _slas,
_vnfImages, _address) public {
    newNetworkProvider(( _name, _cpu, _ram, _storage, _cost, _domain, _slas, _vnfImages,
_address);
    networkProviderIndex = networkProviderIndex + 1;
  }
  function requestResources( _cpu, _ram, uint8 _storage, _domain, _sla, _vnfImage, _time)
public returns {
    uint i = 1;
    uint bestNetworkProvider;
    while( i < networkProviderIndex ) {
      if ( networkProviders[i].cpu >= _cpu && networkProviders[i].ram >= _ram &&
networkProviders[i].storage >= _storage && networkProviders[i].domain == _domain){
        if ( getBestSla(i, _sla) == true && getBestVnfImage(i, _vnfImage) == true) {
          if ( bestNetworkProvider == 0 ) {
            bestNetworkProvider = i;
          else if ( networkProviders[i].cost <= networkProviders[bestNetworkProvider].cost )
            bestNetworkProvider = i;
        }
      i++;
    }
    require(msg.value >= calculateBestCost(bestNetworkProvider, _cpu, _ram, _storage,
_time), "The Ether was not enough (3)");
    uint j = 0;
    while
(providerResourcesTime[ownerToNetworkProvider[msg.sender]][bestNetworkProvider][j] !=
0){
      j++;
    }
    transfer_resources(msg.sender,bestNetworkProvider);
```

Pseudocode 1. Resource management in SC form

### 3.2. Candidate consensus algorithms

One of the main elements of blockchain that has a significant impact on the network's performance is the underlying consensus mechanism. In this study, we focus on two different consensus mechanisms: Raft and IBFT. These two consensus mechanisms were selected because both can be applied in consortium blockchains and perform better (faster block time, higher fault tolerance) than other popular mechanisms, such as PoW and PoS. Moreover, focusing on the impact of consensus, in this solution we decided to maintain the same blockchain characteristics (i.e., number of nodes, blockchain platform, SC structure) in order to facilitate a fair comparison of the consensus mechanisms. On the one hand, Raft [19] is suitable for consortium blockchains where byzantine fault tolerance (BFT) is not a requirement and the key characteristics that should be met is the fast block generation times and the transaction finality. It is worth mentioning that there is no creation of empty blocks in Raft, as it creates blocks on demand. This consensus mechanism is member of the crush fault tolerance algorithms, like Paxos [20], which can guarantee that if a subset of nodes in the decentralized system goes offline the same state of truth is maintained. On the other hand, IBFT [21] consensus mechanism is suitable for private/consortium blockchains where the byzantine fault tolerance is a requirement. This algorithm is member of the BFT consensus family and inherits from the PBFT the 3-phase consensus, PRE-PREPARE, PREPARE and COMMIT. IBFT can tolerate at most F faulty nodes in a N validator network, where N=3F+1. In addition, using this mechanism no forks can be implemented and all valid blocks are appended in the main chain. It should be noted that although IBFT can tolerate the byzantine problem, the block generation times are higher than Raft's because of the use of the 3-phase feature and the BFT characteristic.

The selection between these two consensus algorithms in the considered use case is not easy as there is a tradeoff between security on one hand and speed and fault tolerance of the network on the other. Towards a qualitative comparison, the following should be taken into consideration. NGNs must offer network services to support intensive applications on demand, which means that transactions should be verified very fast, and the block generation time should be low. In addition, the crush fault tolerant attribute is vital for this solution as NGNs support critical applications. Moreover, the use of a consortium blockchain which consists of NPs added by an administration entity, although it reduces the sentiment of decentralization, it also reduces the possibility of the participation of malicious nodes. Furthermore, the identity of each NP is known, which is a fact that discourages NPs from performing malicious actions. As a result, we argue that Raft consensus is more suitable than IBFT in this qualitative approach. However, in the next section the quantitative evaluation of these two consensus algorithms takes place in order to assess if the above rationale is justified by the results and the actual performance limits.

## 4. EVALUATION OF THE SOLUTION

In this section, we aim to evaluate the performance of the blockchain-based marketplace in terms of transaction throughput, latency, and success rate. We first present the testbed that we have set up and then present the results. Our evaluation testbed adopts the architecture described in Figure 1. In this experiment, we deploy the SCs analyzed previously to a Quorum network, and we measure their performance under the two different consensus mechanisms in study.

### 4.1. Description of the testbed and experiment methodology

The experiments are conducted in two different Quorum networks characterized as Systems Under Testing (SUT). The one is using the Raft consensus mechanism and the other IBFT. The Quorum Raft network consists of five nodes in total, where four nodes are hosted in Okeanos cloud while the other one is hosted in an OpenStack infrastructure in UniWA premises (around 500km away) in a dedicated VM. The Quorum IBFT network consists of five nodes as well with the same deployment characteristics. All the VMs that host the blockchain nodes have the same characteristics: 4 vCPUs, 8GB RAM, storage less than 30GB. The operation system was Ubuntu 16.04 and Ubuntu 20.04 respectively.

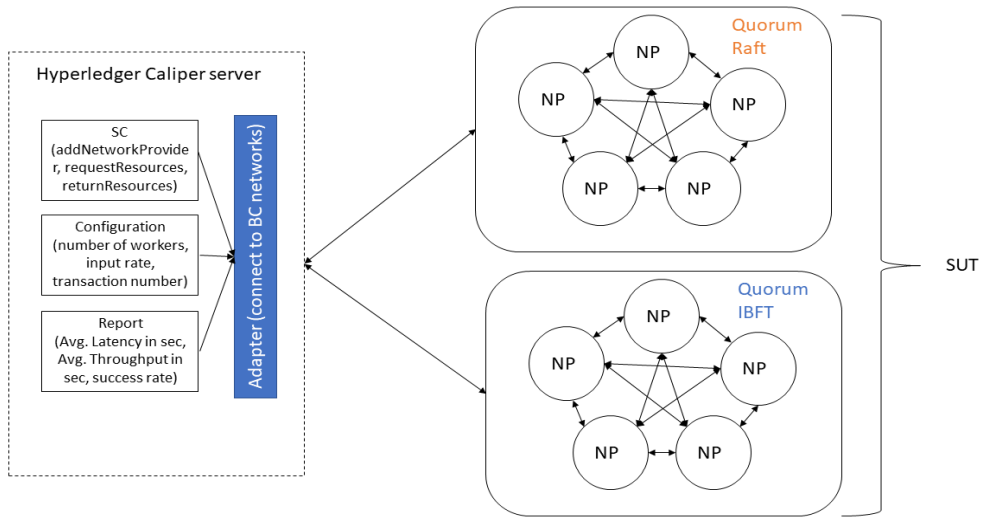

Figure 2. Testbed overview

In both networks, the previously described SC is tested. The blockchain nodes communicate via internet and no internal network is used. The scope of our experiment is to create small, private, geographically distributed blockchain networks which use the Raft and the IBFT consensus respectively and extract useful information in terms of throughput, latency, and success rate. The SC and the behaviour of the blockchain network are tested using the Hyperledger Caliper tool [22], hosted in a VM on Okeanos cloud. Before proceeding to the presentation of the results, we provide clear definitions regarding our metrics because these depend on the tool we used. It is important to stress that there is no other tool that allows for the assessment of such metrics in a real testbed. The metrics we measure are:

• Transaction throughput: The rate at which valid transactions are committed into blocks in the blockchain and become available across all nodes of the blockchain. Throughput is measured in completed transactions per second (TPS) and can be described by the following equation,

$$Transaction\ Throughput = Total\ commited\ transactions/Total\ time\ (s).$$

• Transaction latency: The time elapsed between the submission of a transaction to the time the transaction has been verified and inserted into the blockchain. Once the transaction has been inserted to the blockchain, it is available to all nodes of the network. The transaction latency is measured in seconds and can be described by the following equation:

$$Transaction\ Latency = Confirmation\ time - Submission\ time$$

• Success rate: Once a transaction has been successfully proposed, verified and inserted to a block is considered as a success.

As presented in Figure 2, this tool (Caliper) connects to blockchain networks using a specified adapter compatible with Ethereum and runs tests based on a configuration file created by the user. In our experiment, we focus on the input transaction rate (denoted as ITR). This is the rate at which these transactions are proposed to the blockchain network and is measured in input transaction number per second. Caliper offers controllers which regulate the input transaction pattern. In our experiments we use the fixed rate controller for several ITR values to evaluate the behavior of these two blockchain networks and examine the impact of consensus to the systems' performance. At the end of each experiment round, Caliper produces a report where the Average Transaction Throughput, the Average Transaction Latency and the Success rate are displayed. The same experiments were conducted in both networks and we proceed to the presentation and comparison of the collected results. It should be noted that thanks to Caliper, we managed to check the performance of each function of the SC separately and display the outcome in the following figures.

**4.2. Evaluation results**

We first focus on the success rate for different values of ITR which is depicted in Fig. 3. It is obvious that Raft outperforms IBFT. As we can see, the success rate drops below 100% for IBFT when the ITR is 5 transaction per second while for Raft the success rate remains high for significantly higher ITR values. This was expected due to the crush fault tolerance characteristic of Raft mechanism and the low throughput presented in the IBFT testbed when the ITR increases. Low throughput means low number of transactions that can be validated in a second.

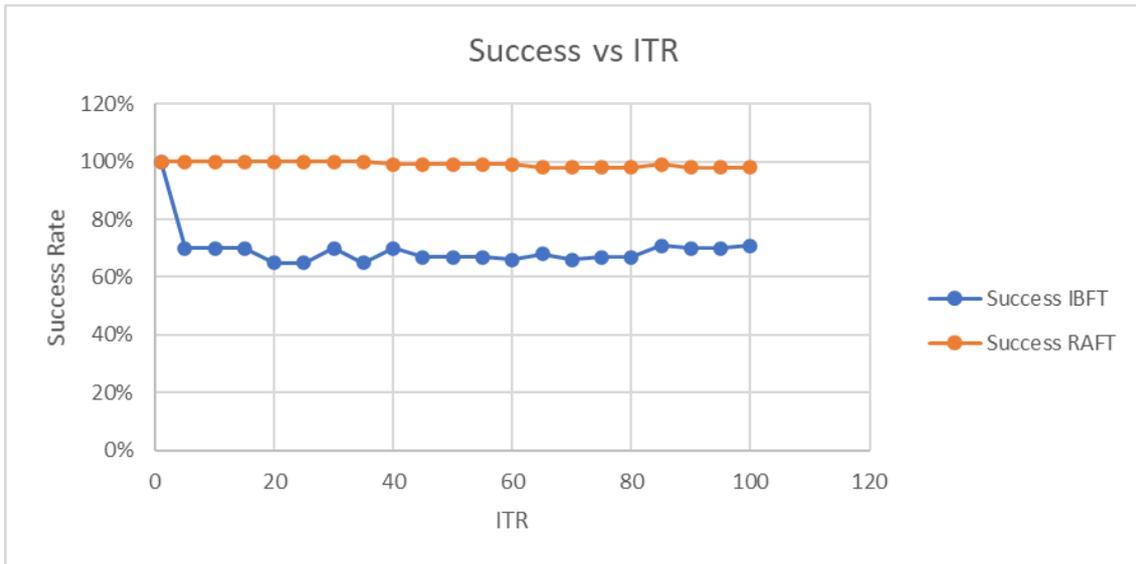

Figure 3. Success rate vs ITR for Raft and IBFT

We then present the throughput rate shown in fig. 4, 5 and 6.

For the selected consensus algorithms we first try to shed light to each of the functions of the SC. The function addNetworkProvider and returnResources behave similarly although the latter presents higher throughput. However, the requestResources function presents a very low throughput as the ITR increases which is caused due to the nature of the function. This function performs recursive queries to find the most suitable NP candidate and select the cheapest one. This process requires more time than the other two functions of the SC. Therefore, we observe high latency in both networks when we focus on the requestResources function. In our experiments we maintain the same ITR for all functions executed in the Raft and IBFT testbed in order to perform a fair comparison. Nevertheless, it is worth mentioning that when we perform an experiment for low ITR value (i.e. ITR = 2) the requestResources function performed better and none failed transaction was presented.

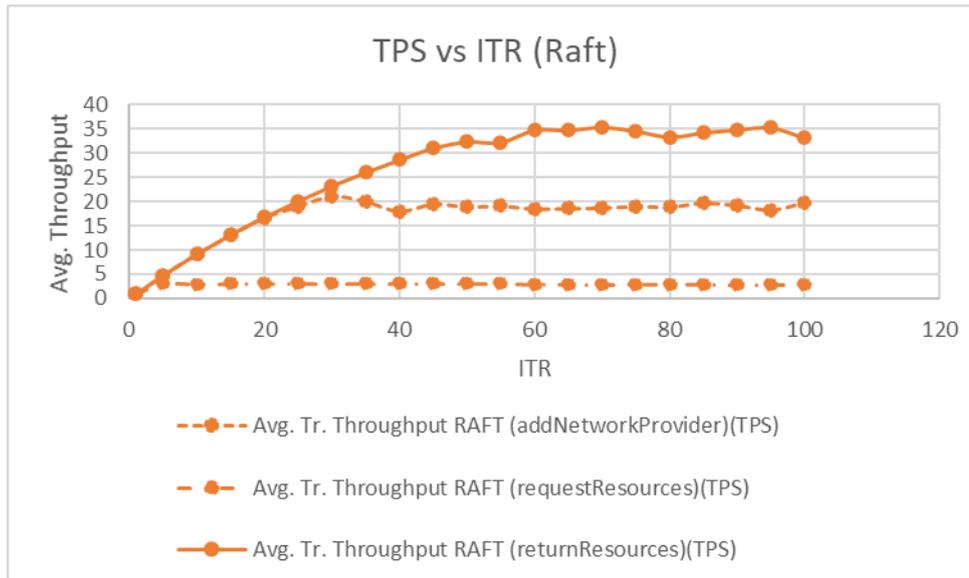

Figure 4. Throughput vs ITR in Raft testbed

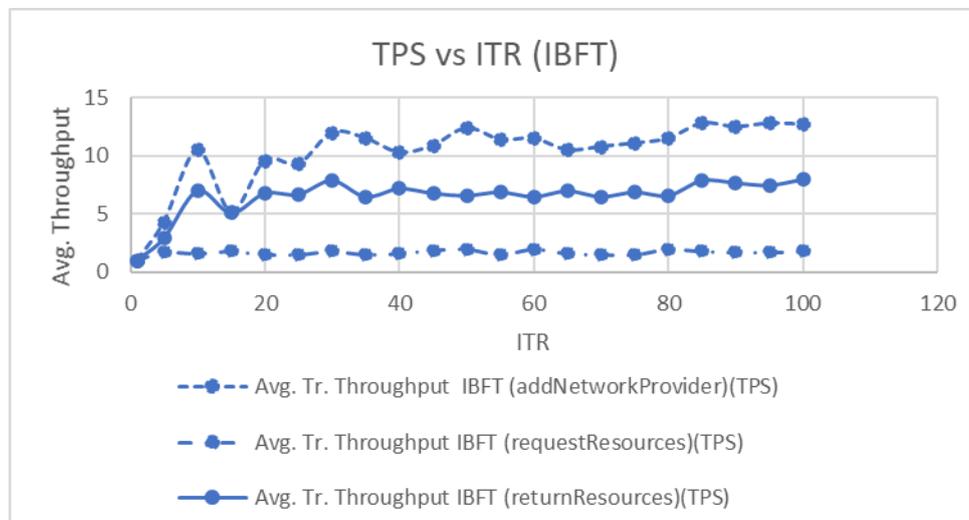

Figure 5. Throughput vs ITR in IBFT testbed

Turning our attention to the comparison of the two consensus algorithms, it is evident that for IBFT the throughput is lower which has caused the high number of losses. For this comparison we take into account the function with the worse performance because this affects the overall performance of the solution. For Raft, the throughput increases with the ITR as expected. From ITR equal to 40 and above, the increase is not linear which indicates that the blockchain network can sustain 40 transaction per second. It is obvious that Raft performs better than IBFT as it presents higher throughput values comparing the performance of the function in each testbed. Figure 6 shows that IBFT is exhibiting half throughput compared to Raft.

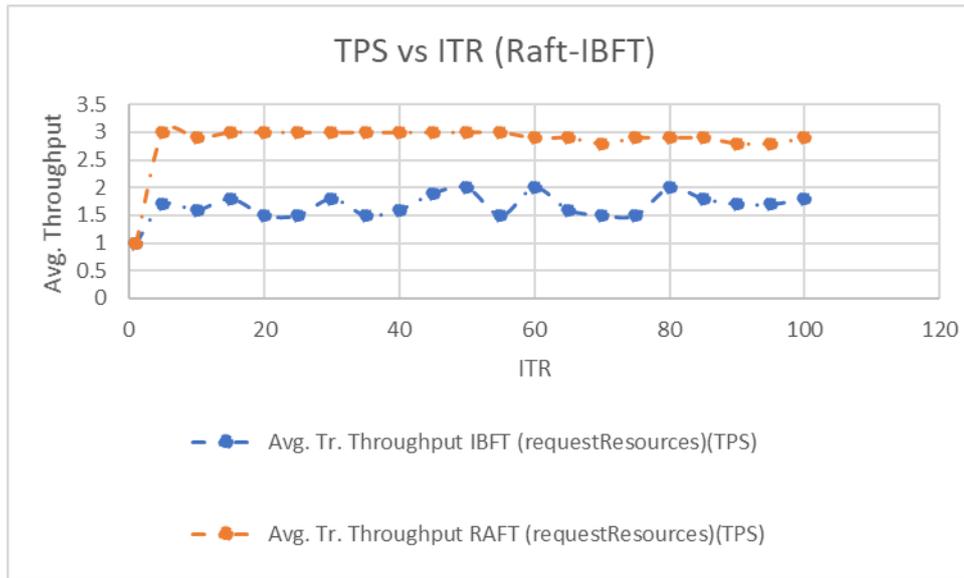

Figure 6. Throughput vs ITR focusing in requestResources function

We finally proceed to the latency which is presented in the following figures for different values of ITR. These figures illustrate the performance of the functions of SC for the two different networks. The blue color corresponds to the IBFT network while the orange represents the Raft. A first observation is that the requestResources function displays higher latency values than any other function of the SC regardless of the consensus used. The functionality of the requestResources described in the pseudocode in the previous sections is more intensive than any other function as it was discussed previously. As such, the latency of this function increases with ITR. The situation is different for the rest two functions which experience almost fixed latency, which depends on the consensus used in the system. IBFT has significant higher latency values than Raft. This was expected due to the characteristics of these two consensus mechanisms described in the previous sections.

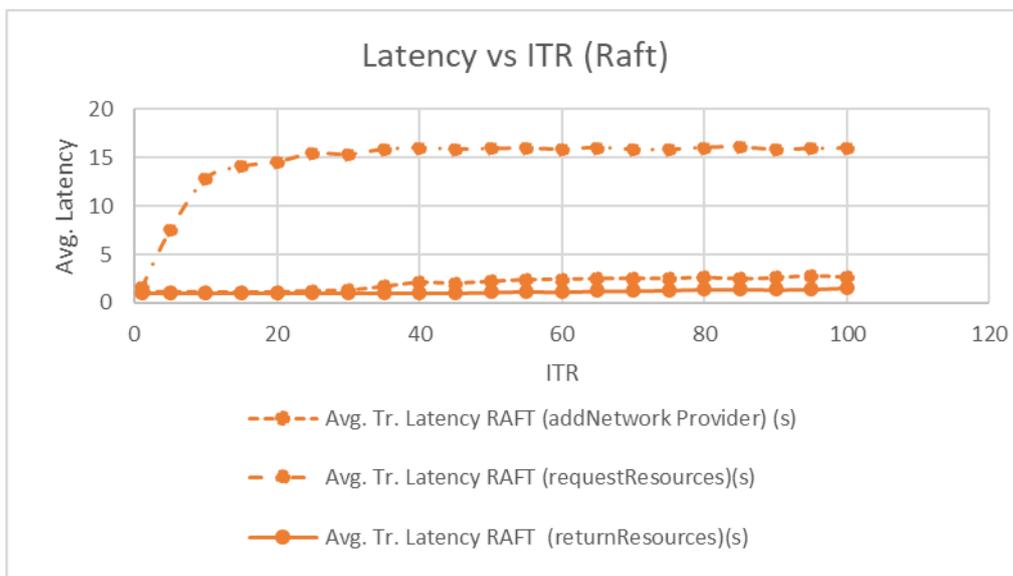

Figure 7. Latency vs ITR in Raft testbed

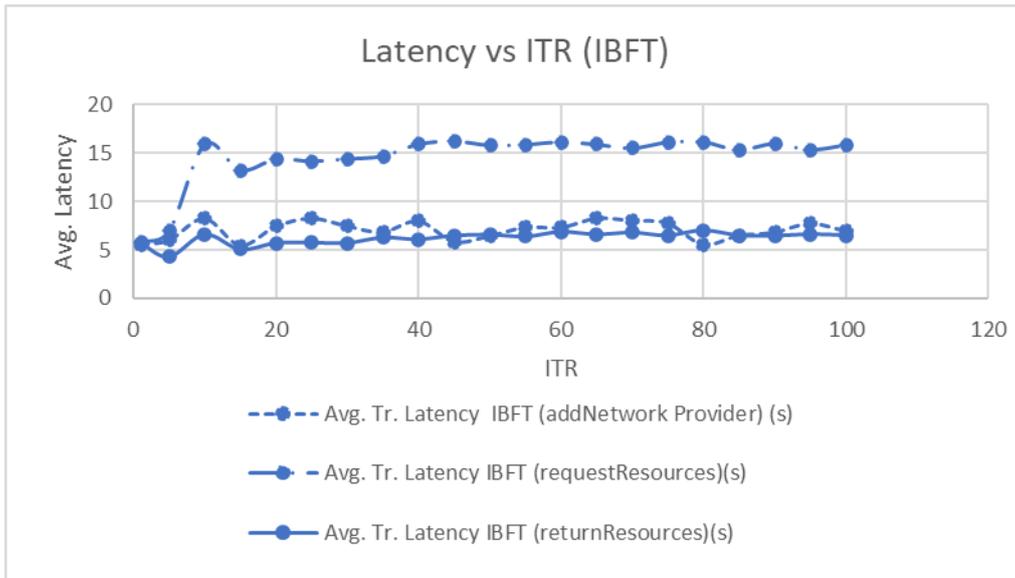

Figure 8. Latency vs ITR in IBFT testbed

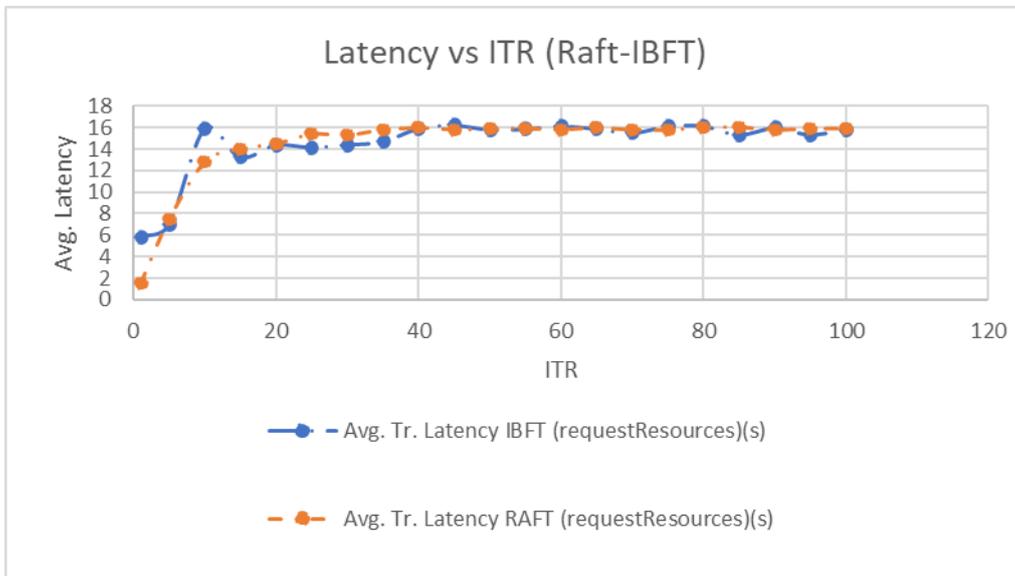

Figure 9. Latency vs ITR in the requestResources function

Raft achieves significantly higher TPS values than IBFT. The behavior of the requestResources function is expected due to the high latency it introduces. The high latency and low throughput values combined with high ITR lead to transaction failures. In Figure 3, IBFT presents lower success rate than Raft. Raft maintains 100% success rate due to the fact that it belongs to the crush fault tolerant consensus family.

The results of the experiment are in line with the theory considering the nature of these two different consensus mechanisms described previously. Consequently, the most suitable consensus among these two is the Raft as it presents better results than IBFT. It is important to be able to compare the latency to the service lifetime so that we can decide whether such a solution can work considering resource negotiation per micro-service or per aggregate. For the two SC functions and for RAFT the latency is in the order to seconds which means that the resource request can be performed on a per service basis in the future. This is not the case for the requestResources function which is about 17s. The latency for both consensus algorithms for this function is almost the same because the processing time dominates the consensus algorithm execution time. Therefore, from the perspective of the solution designer, it is important to either further optimize the code of the SC regarding the request resources transaction function, or to have this function executed outside the blockchain and register in the blockchain only the result of the function.

To improve the situation, the use of oracles and the development of oracle-enabled services is suggested. The oracle acts as a middleware that connects the blockchain world with the outside services in a secure manner. Recently, many oracle mechanisms have been developed [23, 24] and can be characterized based on the data source, the trust model, the design pattern and the interaction with the blockchain [25]. Focusing on the security and integrity of the oracle, the examination of the trust model that should be adopted is crucial. There are two main categories: the centralized trust model (which use a mechanism to prove the authenticity of the data they exchange with the blockchain network) and the decentralized, which use many oracles which in turn use consensus mechanisms to safeguard the interaction with the blockchain. However, although the latter seems to be closer to the nature of blockchain, it adds latency to the system since it introduces an extra consensus mechanism.

## 5. CONCLUSIONS

One of the most interesting research topics is the role of blockchain technology in NGNs and how it can contribute to the evolution of networking sector. The use of blockchain for the efficient resource management in modern NGNs marketplaces has triggered the interest of industry and academia as many works have been published. This paper focuses on this field of study and presents a blockchain-based solution which is implemented in a SC. This SC is deployed in a blockchain environment and presents a resource management scenario in NGNs marketplace. In contrast to other related works, we proceed to the examination and testing of two consensus mechanisms Raft and IBFT to identify which is the most suitable for this use case. The main metrics we focused on are the transaction throughput, transaction latency and the success rate. The experiments were conducted and described in detail in order to justify the selection of the most suitable consensus and check the feasibility of this solution. After the evaluation of the results derived from the experimental process, we proceed to the identification of the points that need to be improved.

NGNs are responsible to provide high quality network services on demand with features such as ultra-low latency and high throughput. Therefore, the resource management process is extremely sensitive to time variations and latency. After the evaluation of our experiments, we may claim that the use of an AI-assisted prediction mechanism could improve the overall performance and increase the feasibility of the solution. This radically changes the scene compared to the traditional resource management which was performed in each network sectors separately, putting emphasis and implementing intelligence in each domain considering the domain resource inelastic [26]. Moreover, the use of oracles in the blockchain can lead to the development of more efficient applications which can interact with many other web services. The introduction of such mechanisms and their impact on the systems' efficiency is a topic that we will examine in our future work as well as, the combination of AI and blockchain in NGNs. In our case, an oracle service could be an AI-assisted prediction mechanism which could be used in this scenario to reduce the overall time needed for a network service to be offered. The idea is the use of a prediction mechanism that will notify NPs about the upcoming network demands. Then NPs could trigger the SC's functions in time and acquire the necessary resources.

**Authors**

**Michael Xevgenis** is a PhD Candidate at the Department of Industrial Design and Production Engineering of the University of West Attica. The subject of his PhD research is the Secure Resource Management in Next Generation Networks, where the Blockchain technology is one of the main fields of this study. In addition, he is a Research Associate at the Computer Network and Services Research Team (CONSERT) of the University of West Attica. His research concerns are on the Networking and Cloud Computing sector. The last few years he has participated in two EU projects: the TRILLION and the STORM of H2020. Finally, he holds the Master in Data and Networking Communications of Kingston University in collaboration with TEI of Piraeus.

**Dimitrios G. Kogias** was born in Athens in 1978. He received his diploma in Physics from the National and Kapodistrian University of Athens in 2001. In December 2004 he received his M.Sc. in Electronics and Radioelectrology and in May 2010 his Ph.D degree from the National and Kapodistrian University of Athens on Algorithms for dissemination of information in Unstructured Networking Environments. He works as an Adjunct Lecturer in the Department of Electrical & Electronics Engineering of the University of West Attica (UniWA).

**Ioannis Christidis** has graduated from the University of West Attica while also participated in ASSET Learning Community & Ecosystem. He is currently a postgraduate student at the University of West Attica and the subject of his research is Blockchain as real time monitoring system. His research interests focuses around blockchain and distributed ledger technologies, data freshness and gamification.

**Charalampos Z. Patrikakis** is a Full Professor at the Dept. of Electrical & Electronics Engineering of University of West Attica. He has participated in more than 32 National, European and International programs, in 16 of which he has been involved as technical coordinator or principal researcher. He has more than 100 publications in chapters of books, international journals and conferences, and has 2 contributions in national legislation. He is a member of the editorial committee of more than 50 international journals and conferences, and has acted as editor in the publication of special issues of international journals, conference proceedings volumes and coedited three books. He is a senior member of IEEE, Assistant Editor In Chief (Special Issues) of IEEE IT Pro Magazine, member of the Technical Chamber of Greece, the European

Association for Theoretical Computer Science, ACM, and counselor of the IEEE student department of University of West Attica.

**Helen C. Leligou** received the Dipl.Ing. and Ph.D. degrees, both in electrical and computer engineering, from the National Technical University of Athens (NTUA), Athens, Greece, in 1995 and 2002, respectively. Her research interests lie in the area of protocol design for communication systems, access control mechanisms in optical access/metro/core networks, with emphasis on their implementation in hardware using state-of-the-art primarily FPGA but also ASIC technologies. She has also been active in security, routing and application layer protocols for wireless sensor networks and their implementation in embedded systems. Currently, her research interests lie in blockchain technologies and in application of novel sensor applications for affect detection in learning environments. She is currently assistant professor at the University of West Attica while from 2007-2017 she is acting as Assistant Professor at Technological Educational Institute of Central Greece teaching "Digital Design" and "Computer Networks". Her research results have been published in more than 100 scientific journals and conferences. She has participated in several EU-funded ACTS, IST and ICT and H2020 research projects in the above areas.